\documentclass[prb,aps,twocolumn]{revtex4}
\usepackage{graphicx,amsfonts,bm,amsmath}

\newcommand{\bkl}{b_{\bm{k},\lambda}}

\newcommand{\bkld}{b_{\bm{k},\lambda}^{\dag}}

\newcommand{\sumkl}{\sum_{\bm{k},\lambda}}
\newcommand{\sumqlp}{\sum_{\bm{q},\lambda'}}
\newcommand{\wk}{\omega_{\bm{k}}}
\newcommand{\wq}{\omega_{\bm{q}}}

\newcommand{\gkl}{g_{\bm{k}\lambda}}
\newcommand{\gqlp}{g_{\bm{q}\lambda'}}
\newcommand{\kk}{\bm{k}}
\newcommand{\qq}{\bm{q}}

\DeclareMathOperator{\re}{Re}
\DeclareMathOperator{\im}{Im}

\begin{document}
\title{Collective fluorescence and decoherence of a few nearly
identical quantum dots} 
\author{Anna Sitek}
\author{Pawe{\l} Machnikowski}
 \email{Pawel.Machnikowski@pwr.wroc.pl}
 \affiliation{Institute of Physics, Wroc{\l}aw University of
Technology, 50-370 Wroc{\l}aw, Poland}

\begin{abstract}
We study the collective interaction of excitons in closely spaced
artificial molecules and
arrays of nearly identical quantum dots with the electromagnetic
modes. We discuss how collective
fluorescence builds up in the presence of a small mismatch of the
transition energy. 
We show that a superradiant state of a single exciton in a
molecule of two dots with realistic energy mismatch
undergoes a two-rate decay.
We analyze also the stability of subdecoherent
states for non-identical systems.
\end{abstract}


\maketitle

\section{Introduction}

Confinement of carriers in semiconductor
quantum dots (QDs) leads to spectrally isolated states which
may be optically controlled at a high level of coherence
\cite{zrenner02,stufler05a}. A single QD offers at most two degrees
of freedom (a biexciton) which may be coherently manipulated by
optical fields in various ways \cite{li03,stufler06}, allowing one to
demonstrate the simplest non-trivial quantum logical operations. 
In order to overcome this two-qubit limitation
one needs to develop manufacturing methods
and control schemes for arrays of two and more QDs. Conditional
control of such systems, indispensable in both classical and quantum
computing schemes, requires interaction between the QDs in the array
\cite{biolatti00,sangu04}, which may be provided, e.g., by dipole
interaction between confined excitons \cite{sangu04,quiroga99}.
Therefore, much experimental effort has been devoted to the
investigation of the coupling between QDs and its
signatures in the optical response and correlation statistics of
quantum dot molecules (QDMs) built of two coupled QDs 
\cite{bayer01,ortner03,unold05,krenner05b,gerardot05}.
It turns out that dephasing of excitons in QDMs differs
considerably from that of individual QDs \cite{borri03}. 

Even without interaction, optical properties of QDMs and QD arrays may
be strongly modified due to collective coupling of sufficiently
closely spaced QDs to the electromagnetic (EM) field. 
These collective effects have been extensively
studied for atomic systems
\cite{dicke54,lehmberg70a,lehmberg70b,nussenzweig73,gross82} where they 
manifest themselves by superradiant emission, i.e., an outburst of
radiation from the excited sample, markedly different from any
exponential decay \cite{skribanovitz73}. 
On the other hand, the collective interaction leads
to the appearance of subradiant states for which the 
probability amplitudes for photon emission interfere destructively,
leading to decoupling from the EM reservoir and to
infinite lifetime. It has been proposed to use these states for
noiseless encoding of quantum information \cite{zanardi97}. For the
analogous problem of coupling to phonon modes of a semiconductor,
``subdecoherent states'' of QD arrays have been suggested as a
possible noiseless implementation
\cite{zanardi98b}. Compared to atomic samples, QDs may be
easier to arrange in a regular array but
the perfectly identical transition energy characteristic of natural
atoms is extremely hard to reach for these artificial systems.

The purpose of this paper is twofold. First, we deal with the general,
theoretical problem of
the stability of collective interaction, including noiseless
encoding, against variations of the transition energies. Second, we
look for clear manifestations of interaction between the QDs in the
experimentally observable
optical properties of the QD arrays, depending on the difference 
(mismatch) of the transition energies of individual QDs. 

Thus, we study the interaction between small,
slightly inhomogeneous arrays of QDs and their
EM environment. The system evolution is described within
the Weisskopf--Wigner approach \cite{weisskopf30}. 
We show how the coherent interaction
is destroyed by growing inhomogeneity of the transition energies in
the regime where the latter is comparable to decay rate. As
important examples, we discuss the buildup of superradiant emission
from a few QDs and the decay of ``subdecoherent'' states built on
non-identical QDs. We show that interaction between dots in a
regular array may, to some extent, stabilize the coherence, in contrast
to randomly distributed atomic systems, where it leads to dephasing
\cite{gross82}. We discuss also how the interplay of the 
transition energy mismatch and interaction strength determines the time
evolution of the luminescence of a more realistic QDM. 
We show that the decay of
luminescence varies from exponential with a single QD rate (for
weakly interacting dots), through nonexponential (for interaction
compared to energy difference), again to exponential with doubled
rate (when interaction energy prevails).

The paper is organized as follows. In Section \ref{sec:system} we
define the model used to describe the system. Section \ref{sec:QDM}
describes the collective flourescence and stability of quantum states
in a molecule built of two dots. Next, in Section \ref{sec:QDA} we
extend the discussion to arrays of four dots. The final discussion and
conclusions are contained in Section \ref{sec:concl}.

\section{The system}
\label{sec:system}

We consider an array of QDs located at points
$\bm{r}_{l}$. We assume that each dot may either be empty or contain
one ground state exciton of fixed polarization with an interband
transition energy $E_{l}$, hence can be described as a two-level
system. The dots interact with transverse EM modes
with frequencies $\wk=c\bm{k}$, where $\bm{k}$ is the wavenumber and
$c$ is the speed of light. 
We will describe the system in the interaction picture with respect to
the Hamiltonian
$H_{0}=E\sum_{j}\hat{n}_{j}+\sumkl\wk\bkld\bkl$ ($\hbar=1$),
where $\bkl,\bkld$ are photon creation and annihilation
operators ($\lambda$ labels polarizations),
$E$ is the average of the energies $E_{j}$, and
$\hat{n}_{j}$ is the occupation operator for the $j$th dot, 
$\hat{n}_{j}=\sigma_{+}^{(j)}\sigma_{-}^{(j)}$, where 
$\sigma_{\pm}^{(j)}=\sigma_{x}^{(j)}\pm i\sigma_{y}^{(j)}$ 
and $\sigma_{x,y}^{(j)}$ are Pauli matrices acting on
the $j$th two-level system.
The Hamiltonian of the system is then $H=H_{\mathrm{X}}+H_{\mathrm{I}}$.
The first component describes the excitons,
\begin{equation}\label{hX}
H_{\mathrm{X}}=\sum_{j}\Delta_{j}\hat{n}_{j}
+\sum_{l\neq j}V_{lj}\sigma_{+}^{(l)}\sigma_{-}^{(j)},
\end{equation}
where $\Delta_{j}=E_{j}-E$ are the energy deviations from the average and
$V_{lj}$ are F\"orster couplings between the QDs\footnote{
For extremely closely spaced dots, the
coupling will be dominated by tunnelling, which has the
same structure as the F\"orster term.},
\begin{displaymath}
V_{lj}=\frac{1}{4\pi\varepsilon_{0}\varepsilon_{\mathrm{r}}
r_{lj}^{3}}\left( 
d^{2}-\frac{3|\bm{d}\cdot \bm{r}_{lj}|^{2}}{r_{lj}^{2}} \right),\quad
\bm{r}_{lj}=\bm{r}_{j}-\bm{r}_{l},
\end{displaymath} 
where $\bm{d}$ is the interband dipole moment (for simplicity equal
for all dots), $\varepsilon_{0}$ is the vacuum dielectric constant,
and $\varepsilon_{\mathrm{r}}$ is the relative dielectric constant of
the semiconductor.
For self-assembled dots, 
typical values for this coupling range from $\mu$eV for the distance
$r_{lj}$ of order of 100 nm to meV for closely stacked dots
separated by $\sim 10$ nm\cite{gerardot05,lovett03b,ahn05}. Another
contribution to the coupling may come from the polariton effect
(coupling to transverse field) \cite{parascandolo05}.

The second term in the Hamiltonian accounts for the interaction with the
EM modes in the dipole approximation 
and rotating wave approximation (RWA)
\begin{equation}\label{hI}
H_{\mathrm{I}} =\sum_{l}\sigma_{-}^{(l)}\sumkl \gkl e^{i(\wk-E)}\bkld
+\mathrm{H.c.},
\end{equation}
with
$\gkl=i\bm{d}\cdot\hat{e}_{\lambda}(\kk)
\sqrt{\wk/(2\varepsilon_{0}\varepsilon_{\mathrm{r}}v)}$,
where $\hat{e}_{\lambda}(\kk)$ are unit polarization vectors
and $v$ is the normalization volume for EM 
modes\footnote{The full minimal-coupling Hamiltonian would yield a
cut-off of $\gkl$ at $\wk\sim c/l$, where $l$ is the QD size. For our
discussion it is only important that this frequency is extremely
high, which justifies the Markov approximation. In this approximation,
only the coupling at frequecies close to $\Delta$ is relevant for the
radiative damping.}. 
The QDs
are placed at distances much smaller than the relevant photon
wavelength so that the spatial dependence of the EM field may be
neglected (the Dicke limit). 
For wide-gap semiconductors with $E\sim 1$ eV, zero-temperature
approximation may be used for any reasonable temperature.

In the present discussion, we disregard the coupling of the carriers
with phonons. Let us note that the quantum confinement of
excitons leads to a separation of at least a few meV between the
ground exciton state involved in our analysis and the lowest excited
state in a single dot. Therefore, no real phonon-induced transitions
may take place in a single dot as long as the temperature is low
enough. 
It has been shown that the combination of dipole interaction 
and phonon coupling may lead to phonon-assisted Coulomb transfer
between the dots \cite{govorov05}, which might be responsible for the
uni-directional transfer observed in the experiment \cite{gerardot05}.
However, the estimated rate reaches its maximum of $\sim 2$ ns
for the energy separation of a few meV and decreases considerably away
from this point \cite{govorov05}. 
Therefore, we neglect this effect in the present
considerations. For 
extremely closely spaced dots, with strongly overlapping carrier wave
functions, phonon-assisted tunneling \cite{lopez05} processes might
also take
place on time scales comparable to those characteristic of the
radiative decay. 
Similar to the phonon-assisted Coulomb transfer, 
such processes would lead to thermalization of the
state of a QDM or QD array which, in general, might suppress the
dynamics described in the following sections. 

Another phonon effect on the exciton state is pure dephasing
\cite{krummheuer02,vagov04}. In QDMs, like in individual QDs, such
processes affect only the first few picoseconds of the optical
response of a QDM \cite{borri03}, while our present discussion is
focused on the radiative decay that develops at much longer 
times. Due to this separation of time scales, the evolution related to
the radiative processes may be discussed separately from this
pure dephasing effect. If the system state is prepared by an
ultrafast pulse, the initial phonon dynamics may result
in a certain reduction of the coherence of the initial state, as we
qualitatively discuss in the following sections.

\section{Quantum dot molecules (2 QDs)}
\label{sec:QDM}

We will start our discussion with quantum dot molecules composed of
two QDs. In the present section, we will first discuss the decay of
sub- and superradiant single exciton states in terms of the formal
quantum fidelity with respect to the unperturbed state and in terms of
the experimentally measurable exciton occupation. Then, we proceed to
the decay of the biexciton state which will be studied again in terms
of fidelity and in terms of the measurable photon emission rate.

\subsection{Single-exciton states}

The RWA Hamiltonian 
conserves the number of excitations (excitons plus photons). 
Let us first consider the initial subradiant state 
$|\psi(0)\rangle=(|01\rangle-|10\rangle)/\sqrt{2}$, where the two-digit
kets denote the occupations of the respective dots. Since there is
only one excitation in this state it may, in general, evolve into
\begin{displaymath}
|\psi(t)\rangle  =  c_{01}(t)|01\rangle
+c_{10}(t)|10\rangle
+\sumkl c_{00\kk\lambda}(t)|00,\kk\lambda\rangle,
\end{displaymath}
where the last ket denotes the state with
no excitons and with one photon in the mode $(\kk,\lambda)$.
The Schr\"odinger equation leads to the system of equations for the
coefficients
\begin{subequations}
\begin{eqnarray}
\label{subrad2a}
i\dot{c}_{01} & = & \Delta c_{01}+Vc_{10}
+\sumkl \gkl^{*}c_{00\kk\lambda}e^{i(E-\wk)t},\\
\label{subrad2b}
i\dot{c}_{10} & = & -\Delta c_{10}+Vc_{01}
+\sumkl \gkl^{*}c_{00\kk\lambda}e^{i(E-\wk)t},\\
\label{subrad2c}
i\dot{c}_{00\kk\lambda} & = & \gkl(c_{01}+c_{10})e^{-i(E-\wk)t},
\end{eqnarray}
\end{subequations}
where $V\equiv V_{12}$ and $\Delta\equiv\Delta_{2}=-\Delta_{1}$.
Following the standard Weisskopf--Wigner procedure \cite{scully97} we
formally integrate Eq.~(\ref{subrad2c}) and substitute to
Eq.~(\ref{subrad2a},b), which yields
\begin{eqnarray*}
\dot{c}_{01,10} & = &\mp i\Delta c_{01}-iVc_{10}\\
&& -\int_{0}^{t}ds \mathcal{R}(s)(c_{01}(t-s)+c_{10}(t-s)),
\end{eqnarray*}
where 
$\mathcal{R}(s)=\sumkl|\gkl|^{2}e^{i(E-\wk)t}$
is the memory function of the photon reservoir. As the latter decays
extremely quickly compared to the timescales of the evolution of
$c_{01,10}$ one can perform the usual Markov approximation and neglect
the $s$ dependence under the integral. Using the fact that 
\begin{displaymath}
\re\int_{0}^{t}ds \mathcal{R}(s)
=\frac{E^{3}|\bm{d}|^{2}}{6\pi\varepsilon_{0}c^{3}}
\equiv\frac{\Gamma}{2},
\end{displaymath}
where $\Gamma$ is the spontaneous decay rate (throughout the paper we
set $1/\Gamma=1$ ns), 
neglecting the imaginary part (i.e., assuming that the Lamb shift
and other radiative corrections are included in the energies), and
defining $\bm{c}=({c_{10},c_{01}})^{T}$
one gets
\begin{equation}\label{hat-A}
\dot{\bm{c}}=\hat{A}\bm{c},\quad
\hat{A}=i\Delta\sigma_{z}-
\left(iV+\frac{\Gamma}{2}\right)\sigma_{x}
-\frac{\Gamma}{2}\mathbb{I},
\end{equation}
where $\sigma_{x,z}$ are Pauli matrices and $\mathbb{I}$ is the unit
matrix.
The reduced density matrix for the charge subsystem may now be easily
constructed as $\rho_{01,01}=|c_{01}|^{2}$,
$\rho_{10,10}=|c_{10}|^{2}$,
$\rho_{01,10}=\rho_{10,01}^{*}=c_{01}^{*}c_{10}$,
$\rho_{00,00}=1-\rho_{01,01}-\rho_{10,10}$, with all the other elements
equal to 0.

\begin{figure}[tb]
\begin{center}
\unitlength 1mm
\begin{picture}(80,30)(0,11)
\put(0,5){\resizebox{80mm}{!}{\includegraphics{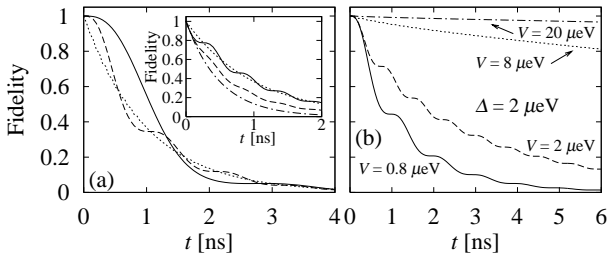}}}
\end{picture}
\end{center}
\caption{\label{fig:fidel2}(a) The fidelity for a
subradiant state for $V=0$ and $\Delta=0.8\;\mu$eV (solid),
$2\;\mu$eV (dashed) and $20\;\mu$eV (dotted).
(b) The fidelity for $V\neq 0$,
$\Delta=2\;\mu$eV. Inset in (a)
shows the fidelity of a superradiant state for different dots without
interaction ($\Delta=4\;\mu$eV, solid) and for interacting dots 
($\Delta=2.68\;\mu$eV, $V=2.97\;\mu$eV, dashed), compared to an exponential decay with
the rates $\Gamma$ and $2\Gamma$ (dotted and dash-dotted, respectively).}
\end{figure}

In order to test the stability of the ideally subradiant state in the
case of non-identical dots we denote by $|\psi(t)\rangle$ the pure state evolving from
$|\psi(0)\rangle$ in the absence of the EM reservoir ($\Gamma=0$) and
define the fidelity of the actual state $\rho$ by 
$F=\sqrt{\langle\psi(t)|\rho|\psi(t)\rangle}$. In
Fig.~\ref{fig:fidel2}(a) we show the result for a few values of the energy
difference $\Delta$ in the limit of vanishing F\"orster coupling
between the dots (i.e., for sufficiently distant dots). It is clear that
the state maintains its subradiant (stable) character until $t\sim
\pi/(2\Delta)$ but then it enters a superradiant phase and the fidelity rapidly decays
below the value corresponding to an exponential (uncorrelated) decay. 
Depending on the value of $\Delta$, a certain number of
oscillations around this uncorrelated decay rate may be observed. In
the limit of large $\Delta$ these oscillations become very fast,
their amplitude decreases, and the decay closely follows that of
uncorrelated systems, as expected for systems with large energy
difference and therefore interacting with disjoint frequency ranges of
the photon reservoir. It is clear
that observing collective effects for such
non-interacting dots requires transition energies identical up to
several $\mu$eV.

If the QDs are close enough, the F\"orster interaction becomes
effective. Since the sub- and superradiant states are eigenstates of the
F\"orster Hamiltonian separated by an energy $2V$, the transition
from the initially subradiant 
state to the superradiant state is suppressed if the magnitude of the
F\"orster coupling exceeds the energy difference
$\Delta$. This is shown in Fig. \ref{fig:fidel2}(b). It is clear that the
decay rate is reduced when $V\sim\Delta$ and the subradiance is
recovered for $V\gg \Delta$. Note that, apart from the trivial
limiting cases, the decay is markedly non-exponential and its
modulation yields information on the origin of the energy level
splitting in the system. Indeed, the decay of the
\textit{superradiant} state
$|\psi(0)\rangle=(|01\rangle+|10\rangle)/\sqrt{2}$ shown in the inset
to Fig. \ref{fig:fidel2}(a) is clearly different for two systems with
the same energy splitting, depending on whether the splitting
originates from the difference between the dots or from the interaction.

\begin{figure}[tb]
\begin{center}
\unitlength 1mm
\begin{picture}(80,30)(0,11)
\put(0,5){\resizebox{80mm}{!}{\includegraphics{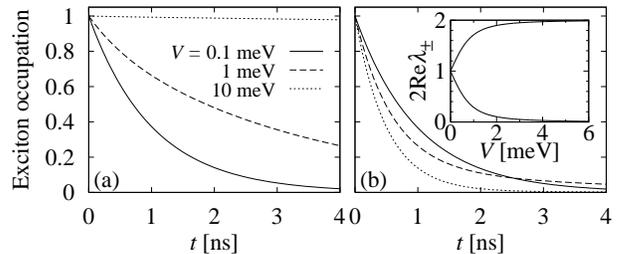}}}
\end{picture}
\end{center}
\caption{\label{fig:occup2meV}The exciton occupation for
sub- (a) and superradiant (b) states for $\Delta=1$ meV. The inset in
(b) shows the values of the occupation decay rates.}
\end{figure}

The signatures of collective interaction with the electromagnetic
field may also be found in the evolution of measurable quantities. As
an example, let us consider the average number of excitons in the
QDM. In the present state of the art of QD manufacturing, the
differences between the transition energies of the two dots are
rather in the meV than in the $\mu$eV range discussed in the previous
case. Therefore, let us consider the evolution of exciton occupations
for the initial states 
$|\psi(0)\rangle=(|01\rangle\pm |10\rangle)/\sqrt{2}$ for a QDM with
$\Delta=1$ meV. The results are shown in
Fig.~\ref{fig:occup2meV}.
As can be seen, in this case the
decay of the occupation shows no oscillations. For
$V\ll\Delta$, both states show simple exponential decay with the rate
$\Gamma$. In the opposite limit, $V\gg\Delta$, the subradiant state
becomes stable while the superradiant state decays exponentially with
a twice larger rate. In the intermediate range of parameters, the
decay is not exponential.

The superradiant state is particularly
relevant for optical experiments since such a bright combination of
single-exciton states is excited by ultrafast optical pulses from the
ground state. Thus, the curves plotted in
Fig.~\ref{fig:occup2meV} directly correspond to the decay of
population after an optical excitation.\footnote{Strictly speaking,
the absence of the biexciton component requires either a large enough
biexcitonic shift or an excitation weak enough to neglect the
higher-order biexcitonic occupation.}
Therefore, let us study this decay in more detail.

The system evolution is governed by the matrix $\hat{A}$
[Eq.~(\ref{hat-A})]
which, for $\Gamma\ll\Delta$, is nearly purely
anti-Hermitian. Therefore, its eigenvectors are nearly orthogonal to
each other. Up
to corrections of order of $\Gamma/\Delta\ll 1$, they can be written in
the form 
\begin{displaymath}
\bm{u}_{+}=\left( \begin{array}{c} \cos\varphi \\ -\sin\varphi
\end{array}\right),\quad
\bm{u}_{-}=\left( \begin{array}{c} \sin\varphi \\ \cos\varphi
\end{array}\right),
\end{displaymath}
where 
\begin{displaymath}
\sin\varphi= \frac{1}{\sqrt{2}}\left[ 
1-\frac{\Delta}{\sqrt{\Delta^{2}+V^{2}}} \right]^{1/2}.
\end{displaymath}
The corresponding eigenvalues are
\begin{displaymath}
\lambda_{\pm}=-\frac{\Gamma}{2}\pm
\sqrt{-\Delta^{2}+(iV+\Gamma/2)^{2}}.
\end{displaymath}
The solution of Eq.~(\ref{hat-A}) for the superradiant initial
condition is
\begin{displaymath}
\bm{c}(t)=\sin(\varphi+\pi/4)\bm{u}_{-}e^{\lambda_{-}t}
+\cos(\varphi+\pi/4)\bm{u}_{+}e^{\lambda_{+}t}.
\end{displaymath}
The number of excitons therefore evolves as 
\begin{eqnarray*}
n(t)=|\bm{c}(t)|^{2} & = & \sin^{2}(\varphi+\pi/4)
e^{2\re\lambda_{-}t}\\
&&+\cos^{2}(\varphi+\pi/4)e^{2\re\lambda_{+}t}.
\end{eqnarray*}
Due to the almost perfect orthogonality of the eigenvectors
$\bm{u}_{\pm}$ the interference term vanishes and the occupation
decay is a combination of two exponentials with different rates, as shown in
Fig.~\ref{fig:occup2meV}(b). In the inset to this figure we show the
values of the two decay constants as a function of $V$ for $\Delta=1$
meV. 

If the initial sub- or superradiant state of a QDM is prepared by an
ultrafast optical pulse it will partly loose its coherence within a
few picoseconds of the system evolution due to phonon-induced pure
dephasing. The detailed dynamics of this dephasing
process differs from that of a single QD and depends on the system
geometry \cite{grodecka06}. Nonetheless, its essential effect is to
perturb the superposition state towards a mixture of two states, each
of which undergoes the usual exponential decay. Therefore, one may
expect a decrease of the amplitude of the oscillations in
Figs.~\ref{fig:fidel2} and \ref{fig:super2} and a shift of the decay
curves in Fig.~\ref{fig:occup2meV} towards the monoexponential decay
with the usual decay rate. For special values of the energy mismatch,
the results may also be modified by the phonon-assisted Coulomb
transfer \cite{govorov05}. 

\subsection{Biexciton state}

Next, let us consider the case of the same two QDs, but initially
excited to the $|11\rangle$ state. This state can be experimentally
prepared in various ways \cite{stufler06,li03}.
The general form of the state is now
\begin{eqnarray*}
\lefteqn{|\psi(t)\rangle = c_{11}(t)|11\rangle}\\
&&+\sumkl c_{01\kk\lambda}(t)|01\kk\lambda\rangle
+\sumkl c_{10\kk\lambda}(t)|10\kk\lambda\rangle\\
&&+\sum_{\kk,\lambda,\qq,\lambda'} c_{00\kk\lambda\qq\lambda'}(t)
|00,\kk\lambda\qq\lambda'\rangle,
\end{eqnarray*}
and the amplitudes evolve according to the equations
\begin{subequations}
\begin{eqnarray}
\label{super2a}
i\dot{c}_{11} & = & 
\sumkl\gkl^{*}
\left(c_{01\kk\lambda}+c_{10\kk\lambda}\right)e^{i(E-\wk)t},\\
\label{super2b}
i\dot{c}_{01\kk\lambda} & = & \Delta c_{01}+Vc_{10}
+\gkl c_{11}e^{-i(E-\wk)t}\\
\nonumber
&&+\sumqlp \gqlp^{*}c_{00\kk\lambda\qq\lambda'}e^{i(E-\wq)t},\\
\label{super2c}
i\dot{c}_{10\kk\lambda} & = & -\Delta c_{10}+Vc_{01}
+\gkl c_{11}e^{-i(E-\wk)t}\\
\nonumber
&&+\sumqlp \gqlp^{*}c_{00\kk\lambda\qq\lambda'}e^{i(E-\wq)t},\\
\label{super2d}
i\dot{c}_{00\kk\lambda\qq\lambda'} & = & 
\gqlp(c_{01\kk\lambda}+c_{10\kk\lambda})e^{-i(E-\wq)t}. 
\end{eqnarray}
\end{subequations}
As previously, we formally integrate Eq.~(\ref{super2d}),
insert it into Eqs.~(\ref{super2b},\ref{super2c}), and use the
short memory assumption. This yields the equation for 
$\bm{c}_{\kk\lambda}=(c_{10\kk\lambda},c_{01\kk\lambda})^{T}$,
\begin{equation}\label{c}
\bm{c}_{\kk\lambda}=-i\gkl\int_{0}^{t}ds e^{\hat{A}(t-s)}c_{11}(s)
e^{-i(E-\wk)s}\bm{b},
\end{equation}
where $\bm{b}=(1,1)^{T}$ and $\hat{A}$ is defined in
Eq.~(\ref{hat-A}). 
Substituting this in turn into Eq.~(\ref{super2a}) we find
\begin{displaymath}
\dot{c}_{11}=-\int_{0}^{t}ds
\mathcal{R}(s)\bm{b}^{T}e^{\hat{A}s}\bm{b}c_{11}(t-s). 
\end{displaymath}
Since the elements of $\hat{A}$ are of order of $\mu$eV or meV, the
matrix exponent is slowly varying on the timescales of reservoir
memory, as is $c_{11}(t)$, and both may be taken at $s=0$, which leads
to the decay equation in the usual form 
$\dot{c}_{11}=-\Gamma c_{11}$ or, for the corresponding element of the
reduced density matrix, $\dot{\rho}_{11,11}=-2\Gamma\rho_{11,11}$.

The evolution equations for the other elements of the density matrix
may be found by writing, for instance,
$\dot\rho_{01,01}=2\re \sumkl (c_{10\kk\lambda}^{*}\dot{c}_{10\kk\lambda})$,
substituting $\dot{c}_{10\kk\lambda}$ from Eq.~(\ref{c}), and using
once more the short memory approximation. Performing this
procedure for all the elements of $\rho$ in the single-exciton sector,
one arrives at the equations
\begin{eqnarray*}
\dot{f}_{11} & = & -2\Gamma f_{11},\\
\dot{f}_{10} & = & -\Gamma(\re p+f_{10}-f_{11})+2V\im p,\\
\dot{f}_{01} & = & -\Gamma(\re p+f_{01}-f_{11})-2V\im p,\\
\dot{p} & = & -2i\Delta
p-\Gamma(p+\frac{f_{01}+f_{10}}{2}-f_{11})\\ 
&& +iV(f_{01}-f_{10}),
\end{eqnarray*}
where we denoted $f_{lj}=\rho_{lj,lj}$, $l,j=0,1$, and
$p=\rho_{01,10}$.

\begin{figure}[tb]
\begin{center}
\unitlength 1mm
\begin{picture}(80,30)(0,11)
\put(0,5){\resizebox{80mm}{!}{\includegraphics{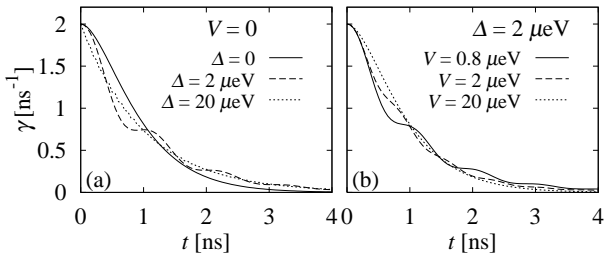}}}
\end{picture}
\end{center}
\caption{\label{fig:super2}Photon emission rate for a
superradiant state for $V=0$ (a) and for $V\neq 0$,
$\Delta=2\;\mu$eV (b).}
\end{figure}

The photon emission rate
$\gamma=-(2\dot{f}_{11}+\dot{f}_{01}+\dot{f}_{10})$ for the initial
state $|11\rangle$ is plotted in Fig.~\ref{fig:super2}. In the case of
$V=0$ [Fig.~\ref{fig:super2}(a)] we see that the photon emission loses
its superradiant behavior for growing energy mismatch between the
dots, tending to the usual exponential decay for large
$\Delta$. Like in the previous case, removing the degeneracy
between the sub- and superradiant single-exciton states by including
the F\"orster coupling stabilizes the collective fluorescence [the
dotted line in Fig.~\ref{fig:super2}(b) coincides with the $\Delta=0$
line in Fig.~\ref{fig:super2}(a)].

\section{Quantum dot arrays (4 QDs)}
\label{sec:QDA}

In this section, we study arrays of four QDs in a very special,
regular arrangement. The resulting symmetry of the F{\"o}rster
term leads to symmetric eigenstates and, as we show below, to the 
stabilization of collective effects.

\begin{figure}[tb]
\begin{center}
\unitlength 1mm
\begin{picture}(80,30)(0,11)
\put(0,5){\resizebox{80mm}{!}{\includegraphics{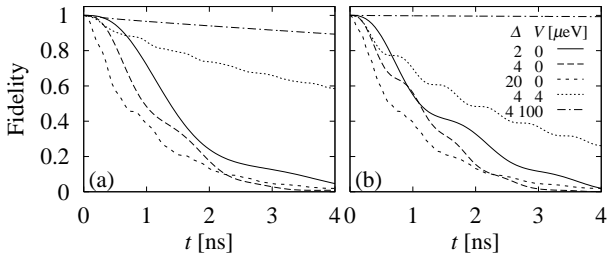}}}
\end{picture}
\end{center}
\caption{\label{fig:fidel4}The fidelity for two
subradiant states $|\psi_{\mathrm{a}\rangle}$ (a) and
$|\psi_{\mathrm{b}\rangle}$ (b) of 4 QDs for various combinations of parameters.
The parameter values given in (b) are
valid for both figures.} 
\end{figure}

In general, the Weisskopf--Wigner equations
lead to the Lindblad equation for the evolution of the reduced density
matrix of the charge subsystem 
\begin{equation}\label{Lind}
\dot{\rho}=-i[H_{\mathrm{X}},\rho]+\mathcal{L}[\rho],
\end{equation}
with 
\begin{displaymath}
\mathcal{L}[\rho]=\Gamma \left[ \Sigma_{-}\rho\Sigma_{+}
-\frac{1}{2}\left\{\Sigma_{+}\Sigma_{-},\rho \right\}_{+}\right],
\end{displaymath}
where $\Sigma_{\pm}=\sum_{j}\sigma_{\pm}^{(j)}$.
We now use Eq.~(\ref{Lind}) to study the evolution of 
four QDs forming a square array in the $xy$
plane. 
The energy deviations of individual dots are now
$\Delta_{i}=\alpha_{i}\Delta$, where $\sum_{i}\alpha_{i}=0$ and
$\sum_{i}\alpha_{i}^{2}=1$, so that
$\Delta$ is the mean square variation of the transition energies. The
details of the system evolution depend on the particular choice of
$\alpha_{i}$ but the general behavior is only governed by the
interplay of $\Delta$ and $V$ (unless some particularly symmetric
choice is made). We arbitrarily fix $\alpha_{1}=0$,
$\alpha_{2}=-0.8$, $\alpha_{3}=0.27$, $\alpha_{4}=0.54$ and use the mean
square variation $\Delta$ as a parameter. The F\"orster
interaction is parameterized by its magnitude $V$, with 
$V_{12}=V_{23}=V_{34}=V_{41}=V$ and $V_{13}=V_{24}=2^{-3/2}V$
(the dots are numbered clock-wise).

First, let us choose the subradiant initial states
$|\psi_{\mathrm{a}}(0)\rangle
=(|1001\rangle-|0101\rangle+|0110\rangle-|1010\rangle)/2$
and 
$|\psi_{\mathrm{b}}(0)\rangle
=(|1001\rangle-|0011\rangle+|0110\rangle-|1100\rangle)/2$, 
which span the subspace of logical qubit states that may be used for
noiseless encoding of quantum information on four physical qubits
\cite{zanardi97}. 
Obviously, for non-identical dots the phases in these superpositions
will rotate and the state will be driven out of the initial noiseless
subspace which leads to a decrease of fidelity, as shown in
Fig.~\ref{fig:fidel4}. 
Out of the two states, only $|\psi_{\mathrm{b}}(0)\rangle$ is a
non-degenerate eigenstate of the Fr\"ohlich interaction for the square
array. As a result, as can be seen in
Fig.~\ref{fig:fidel4}, only this state is
fully stabilized by the F\"orster interaction for $V\gg\Delta$ (the
lines for $V=100\;\mu$eV in Fig. \ref{fig:fidel4} are very close to
the asymptotic case of $V\to\infty$). 
Since the other state
$|\psi_{\mathrm{a}}(0)\rangle$ is never completely stable
the entire ``noiseless subspace'' of logical states \cite{zanardi98b}
remains stable only for an extremely homogeneous array of QDs.

\begin{figure}[tb]
\begin{center}
\unitlength 1mm
\begin{picture}(80,30)(0,11)
\put(0,5){\resizebox{80mm}{!}{\includegraphics{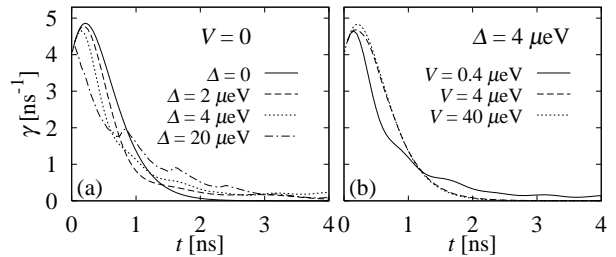}}}
\end{picture}
\end{center}
\caption{\label{fig:super4}Photon emission rate for a
superradiant state of 4 QDs for $V=0$ (a) and for $V\neq
0$ (b).}
\end{figure}

Finally, let us study the photon emission rate from a superradiant
state of four excited QDs, 
$|\psi(0)\rangle=|1111\rangle$ 
(Fig.~\ref{fig:super4}). 
Now, a clear superradiant peak of photon emission develops for
identical dots but vanishes as the dots become different. Again,
interaction between the dots in a regular array stabilizes the
collective emission. It is interesting to note that the
superradiant emission is close to ideal already for $V\sim\Delta$,
while the subradiant states are stabilized only when the interaction
exceeds the energy difference by an order of magnitude.

It should be stressed that the stabilization effect results
from the special, highly symmetric arrangement of the QDs.  
It should be
contrasted with the dephasing induced by analogous interactions in the
randomly distributed atomic samples \cite{gross82}. Likewise, in
an irregular ensemble of QDs obtained by spontaneous self-assembly no
stabilization effect should be expected. 
However, recent progress in the pre-patterned and
strain-engineered growth of QDs \cite{lee01,nakamura03} 
shows great promise for the manufacturing of QD arrays with a desired
geometry. 

\section{Conclusions}
\label{sec:concl}

We have shown that collective interaction
of carries in QDs with their EM environment is
extremely sensitive to the homogeneity of the QD array. Already for
the energy mismatch of order of $\mu$eV the sub- and superradiant
behavior of physical quantities is replaced by their
oscillation around the average exponential decay. Thus, observing
collective fluorescence effects in an ensemble of non-interacting QDs
seems highly unlikely. Likewise, implementing the noiseless encoding
schemes requires the level of homogeneity much beyond the reach of the
present technology.

The destructive effect of inhomogeneity can be, to
some extent, overcome by excitation-transfer coupling (F\"orster or
tunneling) between the dots placed in a regular array. This
can stabilize the subradiance of a state of 2 QDs and the superradiant
emission from 4 QDs but (for a square alignment) still cannot assure
stability of the entire noiseless subspace implemented on 4 QDs.

When the energy mismatch between the dots is of order of meV, like in
the currently fabricated artificial molecules of two QDs, the
oscillations disappear and one observes a decay of the
excitation (exciton occupation number) composed of two exponentials. 
For such a realistic energy
mismatch, the two decay rates for non-interacting dots 
are practically equal to the
free decay rate $\Gamma$. However, with 
growing interaction strength they approach $2\Gamma$ 
(superradiant component) and 0 (subradiant). Thus,
their values carry information on the origin of the energy splitting
(interaction vs. energy mismatch). 


\end{document}